\documentclass[
 reprint,
 amsmath,amssymb,
 aps,prl
]{revtex4-1}

\usepackage{graphicx}
\usepackage{dcolumn}
\usepackage{bm}

\newcommand{\bea}{\begin{eqnarray}}
\newcommand{\eea}{\end{eqnarray}}
\newcommand{\be}{\begin{equation}}
\newcommand{\ee}{\end{equation}}

\begin{document}

\begin{flushright}
CALT-2017-039
\end{flushright}
\title{BPS states, knots and quivers}

\author{Piotr Kucharski$^{1}$, Markus Reineke$^{2}$, Marko Sto$\check{\text{s}}$i$\acute{\text{c}}$$^{3,4}$, and Piotr Su{\l}kowski$^{1,5}$\\
}
\affiliation{\ \\
$^1$ Faculty of Physics, University of Warsaw, ul. Pasteura 5, 02-093 Warsaw, Poland \\
$^2$ Faculty of Mathematics, Ruhr-Universit\"{a}t Bochum, Universit\"{a}tsstrasse 150, 44780 Bochum, Germany \\
$^3$ CAMGSD, Departamento de Matem\'atica, Instituto Superior T\'ecnico,
Av. Rovisco Pais, 1049-001 Lisboa, Portugal   \\
$^4$ Mathematical Institute SANU, Knez Mihailova 36, 11000 Beograd, Serbia \\
$^5$ Walter Burke Institute for Theoretical Physics, California Institute of Technology, Pasadena, CA 91125, USA \\
\
\\
Dedicated to John Schwarz on his 75th birthday
}



\begin{abstract}

We argue how to identify supersymmetric quiver quantum mechanics description of BPS states, which arise in string theory in brane systems representing knots. This leads to a surprising relation between knots and quivers: to a given knot we associate a quiver, so that various types of knot invariants are expressed in terms of characteristics of a moduli space of representations of the corresponding quiver. This statement can be regarded as a novel type of categorification of knot invariants, and among its various consequences we find that Labastida-Mari{\~n}o-Ooguri-Vafa (LMOV) invariants of a knot can be expressed in terms of motivic Donaldson-Thomas invariants of the corresponding quiver; this proves integrality of LMOV invariants, conjectured originally based on string theory and M-theory arguments.

\end{abstract}

\maketitle




In last few decades many intricate links between high energy physics and contemporary mathematics have been found. These links not only helped to solve some specific problems, but also led to discovery of deep and earlier unforeseen relations between different branches of mathematics. In this letter we present a new chain of connections that in a similar vein relates a physical system -- of appropriately engineered branes in string theory -- with mathematical knot theory on one hand, and quiver representation theory on the other hand, thereby revealing deep links between these theories.

It is particularly interesting to determine exact results in string theory. One source of such results are topological string amplitudes. Mathematically they encode Gromov-Witten invariants, and from physics perspective they capture degeneracies of BPS states formed by D-branes. In case of closed string theory these degeneracies are referred to as Gopakumar-Vafa invariants, and are also conjecturally related to Donaldson-Thomas invariants of underlying Calabi-Yau manifolds \cite{Gopakumar:1998jq,maulik2006,Hori:2003ic}. 

Analogous conjectures have been formulated for systems with additional branes. Mathematically they encode open Gromov-Witten invariants, and physically they capture degeneracies of D-branes with boundaries. Furthermore, it has been conjectured that in a system with an appropriate choice of Calabi-Yau manifold and brane configuration, such open amplitudes encode knot invariants, in particular colored HOMFLY-PT polynomials \cite{OoguriV,Labastida:2000zp,Labastida:2000yw}. This relation is a non-trivial consequence of famous relations between knot invariants, Chern-Simons gauge theory, and topological string theory \cite{Witten_Jones,Witten:1992fb}. This relation also predicts, that -- analogously to the closed string case -- colored HOMFLY-PT polynomials are encoded in BPS invariants, referred to as Labastida-Mari\~{n}o-Ooguri-Vafa (LMOV) invariants (or Ooguri-Vafa invariants) \cite{OoguriV,Labastida:2000zp,Labastida:2000yw}. Integrality of these invariants is an important statement for both string theory and knot theory, however it has been verified only in very specific cases e.g. in \cite{OoguriV,Labastida:2000zp,Labastida:2000yw,Ramadevi_Sarkar,Mironov:2017hde}, and more recently for some infinite families of knots and representations \cite{Garoufalidis:2015ewa,Kucharski:2016rlb}. An attempt to prove this conjecture -- which however raised some criticism -- has been made in \cite{Liu:2007kv}.

\begin{figure}[b]
\includegraphics[width=0.45\textwidth]{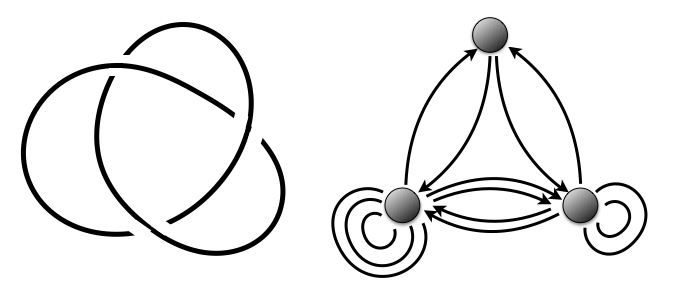} 
\caption{Trefoil knot and the corresponding quiver.}  \label{fig-trefoil}
\end{figure}

By standard arguments \cite{OoguriV,Denef:2002ru,Alim:2011kw}, BPS states in brane systems -- in particular those describing knots -- should have a description in terms of supersymmetric quantum mechanics on their world-volume, which however has not been identified to date. In this letter we argue that such a description actually involves quiver quantum mechanics and identify relevant quivers. To this end we also recall that, in another line of research, it was argued that BPS states in string theory form algebras \cite{Harvey:1996gc}. More recently it has been postulated that such algebras can be identified with Cohomological Hall Algebras \cite{Kontsevich:2010px}, and the associated theory of wall-crossing and motivic Donaldson-Thomas invariants turned out to play an important role in supersymmetric gauge theories and string theory. From mathematical perspective, these ideas can be naturally implemented in the framework of quiver representations and their moduli spaces \cite{Kontsevich:2010px,COM:8276935,Rei12}.

\begin{table}
\begin{tabular}{c|c}
{\bf Knots} & {\bf Quivers} \\
\hline
Homological degrees, framing & Number of loops \\
Colored HOMFLY-PT & Motivic generating series \\
LMOV invariants & Motivic DT-invariants \\
Classical LMOV invariants & Numerical DT-invariants \\
Algebra of BPS states & Cohom. Hall Algebra \\
\end{tabular}
\caption{Identification of quantities associated to knots and quivers.}  \label{tab-duality}
\end{table}

In this letter we argue that BPS states enumerated by motivic Donaldson-Thomas invariants of quiver moduli spaces should be identified with those arising in the supersymmetric quiver quantum mechanics description of brane systems encoding knots. 
This observation has far-reaching consequences: it leads to new ways of interpreting and computing BPS numbers, and to the (idea of the) proof of the famous LMOV conjecture -- and the actual proof for a large class of knots. From mathematical perspective we find direct, unexpected relations between knots and quiver representation theory, which, in particular, lead to a novel categorification of knot invariants. 

In more detail, we claim that to each knot one can associate a quiver, whose moduli space of representations encodes various types of knot invariants, including colored HOMFLY-PT polynomials, homological invariants, etc. In particular LMOV invariants of a given knot can be expressed in terms of motivic Donaldson-Thomas invariants of the corresponding quiver, and integrality of the latter implies integrality of the former ones, which is how the proof of the LMOV conjecture follows.  
Identification of some quantities on both sides of ``knots-quivers'' duality is summarized in table \ref{tab-duality}. As an example, a quiver corresponding to the trefoil knot is shown in fig. \ref{fig-trefoil}.


\subsection{Knot invariants, knot homologies, and LMOV conjecture}

Ever since its birth knot theory attracted attention of physicists, and various knot invariants turned out to have physical interpretation. 
In particular various knot polynomials have been reinterpreted as expectation values in Chern-Simons theory in \cite{Witten_Jones}, and subsequently they were expressed in terms of topological string theory amplitudes \cite{Witten:1992fb}. In \cite{OoguriV} the topological string setup was related to M-theory and it was argued, that colored HOMFLY-PT polynomials are encoded in LMOV invariants, which count bound states of open M2-branes with M5-branes. In this work we show, among others, how -- so far conjectural -- integrality of these invariants can be proven. 

Furthermore, more recently homological knot invariants, such as Khovanov homology and its more involved cousins \cite{Khovanov,KhR1,KhR2}, have been also realized in brane systems in string theory, also revealing yet to be proved new properties of those invariants \cite{GSV,DGR,Gukov:2011ry}. In what follows we show that certain information about homological knot invariants is also encoded in quiver moduli spaces.

Let us introduce first the generating function of $S^r$-colored HOMFLY-PT polynomials, which will be the main object of the subsequent analysis, as well as corresponding LMOV invariants. To start with consider the Ooguri-Vafa generating function \cite{OoguriV,Labastida:2000zp,Labastida:2000yw,Labastida:2001ts}
\be
Z(U,V) = \sum_R  \textrm{Tr}_R U \, \textrm{Tr}_R V = \exp\Big(  \sum_{n=1}^{\infty} \frac{1}{n} \textrm{Tr} U^n \textrm{Tr} V^n \Big),   \label{ZUV}
\ee
where $U=P\,\exp\oint_K A$ is the holonomy of $U(N)$ Chern-Simons gauge field along a knot $K$, $V$ plays a role of a source, and the sum runs over all representations $R$, i.e. all two-dimensional partitions. The LMOV conjecture states that the expectation value of (\ref{ZUV}) takes  form
\begin{align}
\big\langle Z(U,V) \big\rangle &= \sum_R \overline{P}_{R}(a,q) \textrm{Tr}_R V  = \nonumber \\
&= \exp \Big(  \sum_{n=1}^\infty \sum_R \frac{1}{n} f_{R}(a^n,q^n) \textrm{Tr}_R V^n  \Big),    \label{ZUVvev}
\end{align}
where the expectation value of the holonomy is identified with the unreduced HOMFLY-PT polynomial of a knot $K$, $\langle \textrm{Tr}_R U \rangle = \overline{P}_{R}(a,q)=P_R^{\bf 0_1} P_R(a,q)$ (where $P_R^{\bf 0_1}$ is the unknot factor), and the functions 
\be
f_{R}(a,q) = \sum_{i,j} \frac{N_{R,i,j} a^i q^j}{q-q^{-1}}  \label{fR}
\ee
encode conjecturally integer LMOV (or Ooguri-Vafa) invariants $N_{R,i,j}$, counting bound states of M2-branes ending on M5-branes. These functions take form of universal polynomials in colored HOMFLY-PT polynomials.

In what follows we consider one-dimensional source $V=x$. In this case $\textrm{Tr}_R V \neq 0$ only for symmetric representations $R=S^r$, so that $\textrm{Tr}_{S^r}(x) = x^r$. Let us denote $P(x)=\langle Z(U,x) \rangle$, and let $\overline{P}_{r}(a,q)$ denote the $S^r$-colored HOMFLY-PT polynomial of $K$, so that (\ref{ZUVvev}) takes form
\be
P(x) = \sum_{r=0}^\infty \overline{P}_{r}(a,q) x^r =   
e^{\sum_{r,n\geq 1} \frac{1}{n} f_{r}(a^n,q^n)x^{n r}}.
\label{Pz2}
\ee
In this case $f_r(a,q)\equiv f_{S^r}(a,q) = \sum_{i,j} \frac{N_{r,i,j} a^i q^j}{q-q^{-1}}$, for $N_{r,i,j}\equiv N_{S^r,i,j}$, are polynomials in $\overline{P}_{d_1}(a^{d_2},q^{d_2})$ for some $d_1$ and $d_2$ (such that $d_1 d_2\leq r$), e.g. $f_1(a,q) = \overline{P}_1(a,q)$,
\be
f_2(a,q) =  \overline{P}_2(a,q) - \frac{1}{2} \overline{P}_1(a,q)^2 -\frac{1}{2}  \overline{P}_1(a^2,q^2), 
\ee
etc. It follows that (\ref{Pz2}) can be rewritten in product form
\be
P(x) = \prod_{r\geq 1;i,j;k\geq 0} \Big(1 - x^r a^i q^{j+2k+1} \Big)^{N_{r,i,j}} .
\label{Pr-LMOV} 
\ee
Integrality of BPS degeneracies $N_{r,i,j}$ encoded in this product is one important outcome of our work.

Another outcome of our work is a surprising relation of colored HOMFLY-PT polynomials, or their generating function (\ref{Pz2}), to superpolynomials, i.e. the Poincar\'{e} polynomials of the uncolored HOMFLY-PT homologies of knots \cite{DGR}. More generally, one can consider colored HOMFLY-PT homology $\mathcal{H}^{S^r}_{i,j,k}$ that has not been defined rigorously by mathematicians, however its conjectural (reduced) colored superpolynomial
\be
P_r (a,q,t) = \sum_{i,j,k} \, a^i q^j t^k \, \dim \mathcal{H}^{S^r}_{i,j,k}   \label{P_r-aqt}
\ee
can be very effectively computed for various families of knots, using the formalism of differentials \cite{DGR,Rasmussen-differentials,Gukov:2011ry,Gukov:2015gmm}. 
Note that generalizations of the LMOV conjecture to the case of superpolynomials have been considered in \cite{Garoufalidis:2015ewa,Kameyama:2017ryw}.

For example, for trefoil knot $3_1$, eq. (\ref{P_r-aqt}) takes form \cite{FGSS}
\be
P_r(a,q,t)=  \frac{a^{2r}}{q^{2r}}\sum_{k=0}^r {r \brack k} q^{2k(r+1)} t^{2k} \prod_{i=1}^k  (1+a^2q^{2(i-2)}t), \label{Pr-31}
\ee
where ${r \brack k} = \frac{(q^2;q^2)_r}{(q^2;q^2)_k(q^2;q^2)_{r-k}}$, and the $q$-Pochhammer symbol is defined as $(z;q)_n=\prod_{i=0}^{n-1} (1-zq^i)$. For $t=-1$ (\ref{Pr-31}) specializes to the reduced colored HOMFLY-PT polynomial, while in the uncolored ($r=1$) case it reduces to
\be
P_1(a,q,t) = \frac{a^2}{q^2} + a^2 q^2 t^2 + a^4 t^3.   \label{P1-31}
\ee
The monomials in this expression correspond to generators of the HOMFLY-PT homology, and powers of $t$ in each monomial -- in this example taking values $(0,2,3)$ -- are referred to as homological degrees.




\subsection{Moduli of quiver representations}

We now turn to a seemingly unrelated field of moduli of quiver representations. A quiver $Q$ is an oriented graph with a finite set of vertices $Q_0$ and finitely many arrows between vertices $\alpha:i\rightarrow j$. On ${\bf Z}Q_0$, we define the Euler form of $Q$ by
$\langle {\bm d},{\bm e}\rangle_Q=\sum_{i\in Q_0}d_ie_i-\sum_{\alpha:i\rightarrow j}d_ie_j.$
A quiver representation assigns to each vertex $i\in Q_0$ a vector space of dimension $d_i$ and a linear map to each arrow. Recently it turned out that the structure of moduli spaces of quiver representations is very rich, and it is a natural playground for the theory of (motivic) Donaldson-Thomas invariants, Cohomological Hall Algebras, etc.  
In particular various explicit results that we use in what follows are known for symmetric quivers, i.e. such that for any pair of vertices $i$ and $j$, the number of arrows from $i$ to $j$ equals the number of arrows from $j$ to $i$ \cite{Kontsevich:2010px,efimov2012,FR,MR}. 

Of our main focus in what follows will be the following motivic generating series assigned to a symmetric quiver
\be
P_Q(x)=\sum_{{\bm d}\in{\bf N}Q_0}(-q)^{-\langle {\bm d},{\bm d}\rangle_Q} {\bm x}^{\bm d} \prod_{i\in Q_0}\prod_{j=1}^{d_i}\frac{1}{1-q^{-2j}}   \label{PQx}
\ee
where ${\bm x}^{\bm d}=\prod_{i\in Q_0}x_i^{d_i}$. Motivic Donaldson-Thomas invariants $\Omega_{{\bm d},j}\equiv \Omega_{d_1,\ldots,d_m;j}$ (with $m$ denoting the number of vertices) are then defined via the factorization
\be
P_Q(x)=
\prod_{{\bm d}\neq 0} \prod_{j\in\mathbb{Z}} \prod_{k\geq 0} \big(1 - (-1)^j {\bm x}^{\bm d} q^{j+2k+1} \big)^{-\Omega_{{\bm d},j}},   \label{PQx-Omega}
\ee
and proved to be positive integers \cite{efimov2012}. In \cite{MR,FR} two geometric interpretations of coefficients of $\Omega_{\bm d}(q)$ are given: as the intersection Betti numbers of the moduli space of all semisimple representations of $Q$ of dimension vector ${\bm d}$, or as the Chow-Betti numbers of the moduli space of all simple representations of $Q$ of dimension vector ${\bm d}$.


\subsection{Knot invariants from quiver representation theory}

We present now our main claim, which is the statement that various types of knot invariants, for a given knot, are encoded in the data of moduli spaces of quiver representations of a certain quiver, assigned to this knot. As explained in the introduction, from physical perspective this is a consequence of the supersymmetric quiver quantum mechanics description of BPS states in brane systems representing knots. Technically this statement is a consequence of our observation -- which we pose as a general conjecture -- that generating functions of colored HOMFLY-PT polynomials (\ref{Pz2}) can be written in the form
\be
P(x)  = \sum_{d_1,\ldots,d_m\geq 0} q^{\sum_{i,j} C_{i,j} d_i d_j}   \frac{\prod_{i=1}^m x^{d_i} q^{l_i d_i} a^{a_i d_i}(-1)^{t_i d_i}}{\prod_{i=1}^m(q^2;q^2)_{d_i}}  \label{PxC}
\ee
where $C$ is a symmetric $m\times m$ matrix, and $l_i,a_i$ and $t_i$ are fixed integers. Remarkably, this expression has the same form as the motivic generating function (\ref{PQx}) of a certain quiver, up to $q \mapsto -q$ and the specialization 
\be
x_i = x a^{a_i} q^{l_i - 1} (-1)^{t_i}.
\ee
In particular terms proportional to $x^r$ in (\ref{PxC}), with fixed $r$, arise from sets of $\{d_i \}$ such that $r=d_1+\ldots +d_m$. Rewriting of (\ref{Pz2}) in the form (\ref{PxC})  means that the matrix $C$ can be identified as a matrix representing a quiver with $m$ vertices, such that $C_{i,j}$ denotes the number of arrows from vertex $i$ to $j$ (and so $C_{i,i}$ denotes the number of loops at vertex $i$). Therefore, if colored HOMFLY-PT polynomials for a given knot are known, after rewriting the generating function (\ref{Pz2}) in the form (\ref{PxC}), from the form of $C$ the corresponding quiver can be identified. This also means that colored HOMFLY-PT polynomials for a given knot are encoded in a finite number of parameters that determine (\ref{PxC}): the matrix $C$ and integers $l_i,a_i$ and $t_i$. Furthermore, these parameters take specific values, which are encoded in uncolored superpolynomial. Recall that the uncolored, reduced superpolynomial for a given knot (such as (\ref{P1-31}) for the trefoil), is a sum of monomials of the form $a^{a_i} q^{q_i} t^{t_i}$, which correspond to generators of the HOMFLY-PT homology. We claim that the size of the matrix $C$ (and so the number of vertices in the corresponding quiver) is equal to the number of such generators. Moreover, with appropriate ordering of vertices, $t_i$ in (\ref{PxC}) are to be identified with homological degrees of generators of HOMFLY-PT homology, diagonal elements of $C$ are also equal to homological degrees, i.e. $C_{i,i}=t_i$, coefficients of linear powers of $q$ take form $l_i=q_i-t_i$, and $a_i$ are equal to $a$-degree of generators of uncolored HOMFLY homology. An additional minus sign in (\ref{PxC}) comes with the power determined by $t_i$, so that it is relevant only for the generators with odd $t$-grading. Note that it follows, that homological degrees $t_i$ can be read off from the generating series (\ref{Pz2}) rewritten in the quiver-like form, i.e. they are given by the numbers of loops in the corresponding quiver; this means, that the uncolored superpolynomial is encoded in the form of colored HOMFLY-PT polynomials, which is a rather non-trivial and previously unknown statement. 

Note that the values of parameters $l_i, a_i$ and $C_{i,i}$ depend in fact on a choice of normalization of $\overline{P}_r(a,q)$. The values mentioned above, related to unreduced and uncolored superpolynomial, arise when normalization includes only the denominator of the colored HOMFLY-PT polynomial of the unknot $\overline{P}_r(a,q)=P_r(a,q) / (q^2;q^2)_r$. Perhaps more familiar normalization by the full unknot polynomial $\overline{P}_r(a,q)=a^{-r}q^r\frac{(a^2;q^2)_r}{(q^2;q^2)_r}P_r(a,q)$ leads to identification with another version of HOMFLY-PT homology and to twice larger quiver \cite{KRSS2}.

The above conjecture has very nontrivial consequences. In general it implies that various knot invariants are specializations of certain quiver moduli invariants. In particular, under the above specialization the product decomposition (\ref{PQx-Omega}) is identified with the product decomposition (\ref{Pr-LMOV}). It follows that LMOV invariants $N_{r,i,j}$ can be expressed as linear combinations (with integer coefficients) of motivic Donaldson-Thomas invariants $\Omega_{{\bm d},j}\equiv \Omega_{d_1,\ldots,d_m;j}$. As motivic Donaldson-Thomas invariants are proven to be integer, it follows that corresponding LMOV invariants are also integer, which proves the LMOV conjecture! 
We are able to prove this statement for many specific knots (and arbitrary symmetric representations), including some infinite families of knots (e.g. twist knots, certain classes of torus knots), analogously as in the case of trefoil knot, which is analyzed in detail below. Moreover, the limit $q\to 1$ of the motivic generating series immediately implies integrality of classical LMOV invariants $b_{r,i}=\sum_j N_{r,i,j}$ (considered for example in \cite{Garoufalidis:2015ewa}), which are then expressed in terms of (integer) numerical Donaldson-Thomas invariants. Moreover, the fact that LMOV invariants and (generating functions of) colored HOMFLY-PT polynomials are expressed in terms of motivic Donaldson-Thomas invariants -- which arise as certain Betti numbers of quiver moduli spaces -- provides a novel categorification of these knot invariants. 

Some other relations between knots and quivers that follow from our conjecture are listed in table \ref{tab-duality}. For example, the framing by $f\in \mathbb{Z}$ changes the colored HOMFLY-PT polynomial by a factor, that for $S^r$-symmetric representation takes form $a^{2fr}q^{r(r-1)}$. The term with quadratic power of $q$, i.e. $q^{fr^2}=q^{f(\sum_i d_i)^2} = q^{f\sum_{i,j} d_i d_j}$, shifts all elements of $C$ by $f$, which in the dual quiver adds $f$ loops at each vertex and $f$ pairs of oppositely-oriented arrows between all pairs of vertices. Moreover, as predicted in \cite{Kontsevich:2010px}, the Cohomological Hall Algebra associated to a quiver should be identified as the algebra of BPS states \cite{Harvey:1996gc}, which deserves further studies in the context of brane realization of knot invariants.


\subsection{Example -- trefoil knot}

We illustrate the above correspondence in the example of trefoil knot, whose reduced colored HOMFLY-PT polynomials arise by setting $t=-1$ in (\ref{Pr-31}). Using the $q$-binomial identity $(z;q)_k=\sum_i {k \brack i} (-z)^i q^{\frac{i(i-1)}{2}}$, the $q$-binomial and the last product in (\ref{Pr-31}) take form
$$
{r \brack k} \big(\frac{a^2}{q^2};q^2\big)_k = \sum_{i=0}^k \frac{(q^2;q^2)_r  \big( - \frac{a^2}{q^2} \big)^i q^{i(i-1)} }{(q^2;q^2)_{r-k} (q^2;q^2)_i (q^2;q^2)_{k-i}}.
$$
Introducing $r=d_1+d_2+d_3, k=d_2+d_3, i=d_3$ with $d_i\geq 0$, and normalizing $P_r(a,q)$ by $(q^2;q^2)_r$, the generating function (\ref{Pz2}) can be rewritten as in (\ref{PxC})
\begin{align}
&P(x) = \sum_{r=0}^{\infty} \frac{P_r(a,q)}{(q^2;q^2)_r} x^r = \nonumber \\
&\sum_{d_1,d_2,d_3\geq 0} \frac{q^{\sum_{i,j} C_{i,j} d_i d_j  -2d_1-3d_3} (-1)^{d_3} a^{2d_1+2d_2+4d_3} x^{\sum_i d_i}  }{(q^2;q^2)_{d_1} (q^2;q^2)_{d_2} (q^2;q^2)_{d_3}}   \label{PxC-31}
\end{align}
with $
C = 
\left[
\begin{tabular}{ccc}
0 & 1 & 1 \\
1 & 2 & 2 \\
1 & 2 & 3 \\
\end{tabular}
\right]$. The corresponding quiver is shown in fig. \ref{fig-trefoil} and, as we claimed above, its vertices correspond to generators of HOMFLY-PT homology. Furthermore, diagonal elements $(0,2,3)$ of matrix $C$ (representing numbers of loops at vertices of the quiver) indeed agree with homological degrees encoded in the uncolored superpolynomial (\ref{P1-31}), coefficients $l_i=-2,0,3$ of linear terms in $d_i$ in the power of $q$ in (\ref{PxC-31}) are given by $l_i=q_i-t_i$, coefficients $a_i=2,2,4$ in the power of $a$ agree with $a$-degrees of generators of HOMFLY-PT homology, and additional minus sign $(-1)^{d_3}$ is determined by just one generator with odd $t$-degree $t_3=3$.

We stress that rewriting of the generating function of colored HOMFLY-PT polynomials in the form (\ref{PxC-31}) guarantees that it can be written in the product form (\ref{PQx-Omega}), so that LMOV invariants are expressed as combinations (with integer coefficients) of integer motivic Donaldson-Thomas invariants $\Omega_{d_1,d_2,d_3;j}$. This proves the LMOV conjecture for $3_1$ knot, for all symmetric representations.

Analogous calculations can be performed for other knots, e.g. rewriting formulas for colored superpolynomials in \cite{FGSS}. A general formalism facilitating such computations and their consequences is presented in \cite{KRSS2}.


\subsection{Summary}

In this letter we presented a surprising duality between knots and quivers, which underlies the supersymmetric quiver quantum mechanics description of BPS states in brane systems describing knots. To conclude, let us indicate various directions in which relations presented above could be generalized. It is natural to expect that various properties, well known on one side of the duality, should have their counterparts on the other side. 

First, it is desirable to be able to identify a quiver corresponding to a given knot more directly, not necessarily taking advantage of colored HOMFLY-PT polynomials, on which the method presented above relies.

Second, colored HOMFLY-PT polynomials and LMOV invariants are naturally defined for arbitrary representations, while for symmetric representations they are related to algebraic curves of $A$-polynomial type, as well as recursion relations encoded in $\widehat{A}$-polynomials \cite{AVqdef,Garoufalidis:2015ewa,Kucharski:2016rlb}. We expect that analogous objects should exist for quivers, presumably generalizing observations in \cite{COM:8276935}. 

On the other hand, various properties of quivers, their moduli spaces, and other related objects -- in particular Cohomological Hall Algebra -- should shed new light on properties of knot invariants, in particular knot homologies. Further, from the quiver representation viewpoint it is natural to split the generating parameter $x$ into parameters $x_1,\ldots,x_m$, which implies a nontrivial refinement of LMOV invariants (with a label $d$ split into $d_1,\ldots,d_m$), as in (\ref{PQx-Omega}), as well as of HOMFLY-PT polynomials. 

It is also interesting to relate our results to other developments, e.g. to the connection -- albeit in a different context -- between (uncolored) HOMFLY-PT polynomials and Donaldson-Thomas invariants in \cite{Diaconescu-HOMFLY-DT}. Also, the generating functions (\ref{PxC}) take form of combinations of $q$-series that appear in the Nahm's conjecture \cite{Nahm}, which indicates their relation to integrable systems and CFT. 

We believe that the results and conjectures presented in this letter are just a tip of the iceberg, which is worth thorough exploration.


\medskip

{\bf Acknowledgments.}
We thank Sergei Gukov, Satoshi Nawata, Mi{\l}osz Panfil, Yan Soibelman, Richard Thomas, Cumrun Vafa, and Paul Wedrich for discussions and comments on the manuscript. This work is supported by the ERC Starting Grant no. 335739 \emph{``Quantum fields and knot homologies''} funded by the European Research Council under the European Union’s Seventh Framework Programme, and the Foundation for Polish Science. M.S. is partially supported by the Ministry of Science of Serbia, project no. 174012.







\bibliography{abmodel} 

\end{document}